\documentclass[letterpaper,english,reprint,aps,prl,superscriptaddress,showpacs,longbibliography]{revtex4-1}
\usepackage[latin9]{inputenc}
\usepackage{color}
\usepackage{amsmath}
\usepackage{amssymb}
\usepackage{graphicx}
\usepackage{wasysym}

\makeatletter

\pdfpageheight\paperheight
\pdfpagewidth\paperwidth

\newcommand{\lyxdot}{.}

\usepackage{babel}
\usepackage[colorlinks,
            linkcolor=red,
            anchorcolor=black,
            citecolor=blue
            ]{hyperref}

\makeatother

\usepackage{babel}
\begin{document}
\title{Size-Reduction of Rydberg collective excited states in cold atomic
system}
\author{Dong-Sheng Ding}
\email{dds@ustc.edu.cn}

\affiliation{Key Laboratory of Quantum Information, University of Science and Technology
of China, Hefei, Anhui 230026, China.}
\affiliation{Synergetic Innovation Center of Quantum Information and Quantum Physics,
University of Science and Technology of China, Hefei, Anhui 230026,
China.}
\author{Yi-Chen Yu}
\affiliation{Key Laboratory of Quantum Information, University of Science and Technology
of China, Hefei, Anhui 230026, China.}
\affiliation{Synergetic Innovation Center of Quantum Information and Quantum Physics,
University of Science and Technology of China, Hefei, Anhui 230026,
China.}
\author{Zong-Kai Liu}
\affiliation{Key Laboratory of Quantum Information, University of Science and Technology
of China, Hefei, Anhui 230026, China.}
\affiliation{Synergetic Innovation Center of Quantum Information and Quantum Physics,
University of Science and Technology of China, Hefei, Anhui 230026,
China.}
\author{Bao-Sen Shi}
\email{drshi@ustc.edu.cn}
\affiliation{Key Laboratory of Quantum Information, University of Science and Technology
of China, Hefei, Anhui 230026, China.}
\affiliation{Synergetic Innovation Center of Quantum Information and Quantum Physics,
University of Science and Technology of China, Hefei, Anhui 230026,
China.}

\author{Guang-Can Guo}
\affiliation{Key Laboratory of Quantum Information, University of Science and Technology
of China, Hefei, Anhui 230026, China.}
\affiliation{Synergetic Innovation Center of Quantum Information and Quantum Physics,
University of Science and Technology of China, Hefei, Anhui 230026,
China.}
\date{\today}

\maketitle
\textbf{
The collective effect of large amounts of atoms exhibit an enhanced interaction between light and atoms. This holds great interest in quantum optics, and quantum information. When a collective excited state of a group of atoms during Rabi oscillation is varying, the oscillation exhibits rich dynamics. Here, we experimentally observe a size-reduction effect of the Rydberg collective state during Rabi oscillation in cold atomic dilute gases. The Rydberg collective state was first created by the Rydberg quantum memory, and we observed a decreased oscillation frequency effect by measuring the time traces of the retrieved light field amplitude, which exhibited chirped characteristics. This is caused by the simultaneous decay to the overall ground state and the overall loss of atoms. The observed oscillations are dependent on the effective Rabi frequency and detuning of the coupling laser, and the dephasing from
inhomogeneous broadening. The reported results show the potential prospects of studying the dynamics of the collective effect of a large amount of atoms and manipulating a single-photon wave-packet based on the interaction
between light and Rydberg atoms.}

The interaction of light with an atomic media containing a large number
of particles causes a collective effect \cite{W2016Light},
which is at the focus of intense research in different areas in quantum
metrology, quantum optics, and quantum information. The single collective
excitation shared among a large number of ground-state atoms results
in a coherent superposition state \cite{dicke1954coherence,honer2011artificial}.
In contrast to the single atom coupled to the light field this state can
still carry only a single excitation; however, the light matter interaction
is enhanced owing to a large number of ground state atoms, as predicted
by Dicke's theory. When the single excitation
corresponds to the Rydberg excitation, Rydberg-state
super atoms were formed \cite{gaetan2009observation,urban2009observation,dudin2012observation,zeiher2015microscopic,weber2015mesoscopic,beterov2016simulated,paris2017free,busche2017contactless}
consisting of a single Rydberg excitation and many ground state atoms.
Because of the exaggerated properties of the Rydberg atom, it is a valuable
resource for numerous potential applications in quantum computing
\cite{lukin2001dipole,saffman2016quantum}, quantum optics \cite{firstenberg2016nonlinear},
and many-body physics \cite{schauss2012observation,labuhn2016tunable,ding2020phase,ding2021epidemic}
etc.

Storing photons to Rydberg super atoms can be realized
by the technology of quantum memory \cite{bussieres2013prospective,Ding2016}\textcolor{black}{,
in which an interface between light and Rydberg atoms is created that
allows for the storage and retrieval of the optical field.} Demonstrating
a Rydberg-mediated quantum memory could enable the
implementation of quantum computation and information processing with
the advantages of Rydberg super atoms, for example, by converting
a Rydberg super atom to a single photon, the demonstration
of a deterministic single-photon generator can be realized \cite{dudin2012strongly,ripka2018room}.
Coherently preparing and manipulating the Rydberg super atom based on quantum
memory holds promise in quantum information science; it should be studied. When the stored collective excited state is driven by the read field, a Rabi oscillation dynamics of the quantum reading process as studied in a double-$\varLambda$ system \cite{du2008narrowband,mendes2013dynamics,de2014single}, however, an anomalous reduction-frequency oscillation
with a varying frequency has never been reported before.

\begin{figure*}[t]
\includegraphics[width=2\columnwidth]{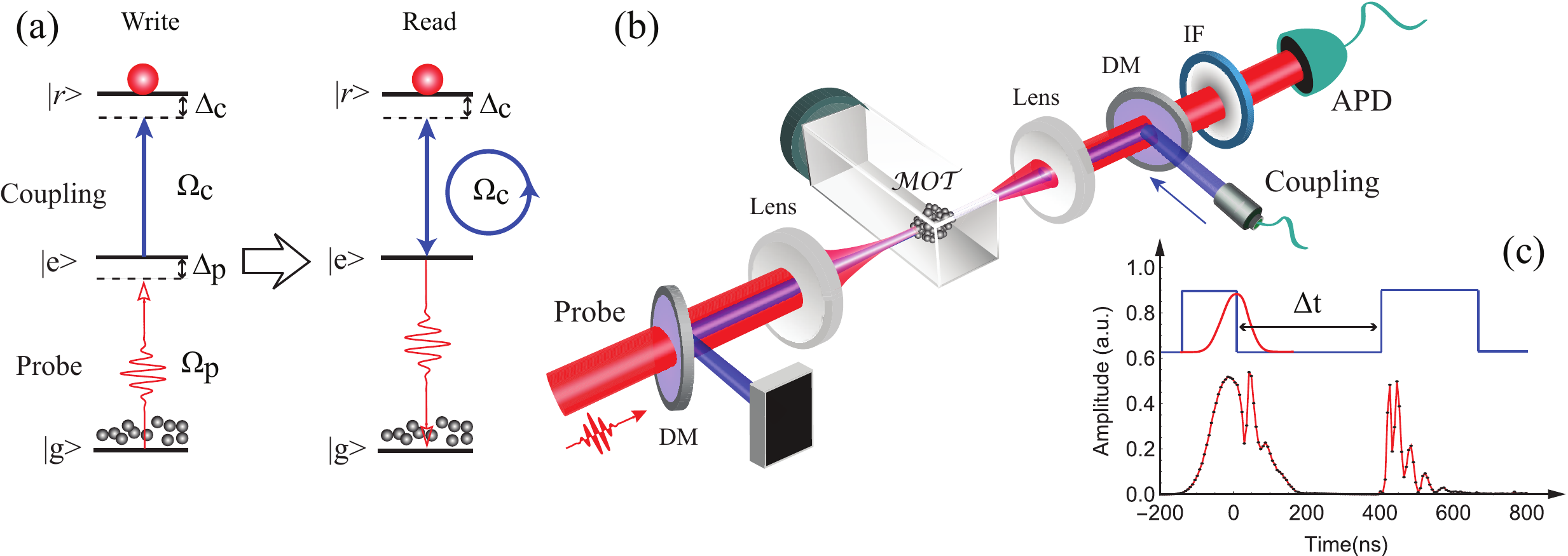}

\caption{\textbf{Experimental setup and diagram.} (a) Schematic of
the energy diagram of write and read processes, the ladder-type atomic
configuration with ground state $5S_{1/2}(F=3)$ ($\left|g\right\rangle $),
excited state $5P_{3/2}(F'=4)$ ($\left|e\right\rangle $) \textcolor{black}{with
a linewidth of $\gamma_{e}$, and highly-excited state $\left|nD_{5/2}\right\rangle $
($\left|r\right\rangle $) with a linewidth of $\gamma_{r}$. $\Omega_{p}$
and $\Omega_{c}$ are the corresponding Rabi frequency of the probe and
coupling fields. The probe (coupling) detuning are denoted by $\Delta_{p(c)}$. The wavelengths of the probe and coupling fields are 780 nm and 480 nm respectively.} (b) Experimental setup of Rydberg
quantum memory. (c) Time sequence for storage process. $\Delta t$
represents the storage time. Labels: DM-dichroic mirror, IF-interference
filter, APD single-photon detector, and MOT magneto-optic traps.}

\label{experimental setup}
\end{figure*}

In this work, we prepared a Rydberg super atom through
quantum memory in the Rydberg electromagnetically induced transparency
(Rydberg EIT) configuration. Rabi oscillation between the low-lying
collective excited-state and high-lying Rydberg-state super atom is
realized by driving the coupling laser in the reading process. The
retrieved probe pulse exhibits chirped characteristics because of
the reduction of the effective size of the Rydberg
super atom. Combining the two-level atoms dephasing
model extracted from inhomogeneous broadening, we model our experimental
observations with a decreased-frequency Rabi oscillation function.
The coherent Rabi oscillation in the Rydberg quantum memory process is
a new representation for combining the collective dynamical behavior of
Rydberg atoms and the radiation of a single photon, which is crucial
for the applications of Rydberg atoms in quantum information processing
\cite{saffman2010quantum} and for providing a versatile interface
between light and atoms.

\textbf{Experimental setup}

The schematics of the energy levels, experimental setup, and time
sequence are shown in figure~\ref{experimental setup}(a)\textendash (c).
The sample media is an optically thick atomic ensemble of Rubidium
85 trapped in MOT. This atomic cloud has a size of $500~\mu$m with
a temperature $\sim$ 20~$\mu$K and an average density of $\sim1.0\times10^{11}\textrm{c\ensuremath{\textrm{m}^{-3}}}$
at the center of the cloud. The optical depth (OD)
in MOT is approximately $20$. The probe field is then input into the atomic
cloud using a beam waist $\sim5$~$\mu\textrm{m}$ in the center of the
MOT estimated by fluorescence imaging, which is covered by the coupling
beam with a beam waist of 16 $\mu\textrm{m}$. With a coupling laser
beam, we demonstrate the quantum memory via Rydberg-EIT in the ladder-type
atomic configuration, consisting of a ground state $\left|g\right\rangle $,
an excited state $\left|e\right\rangle $, and a highly-excited Rydberg
state $\left|r\right\rangle $; here, $n=50$. The probe and coupling
fields are counter-propagating, and couple the two-photon transitions
$\left|g\right\rangle \rightarrow\left|e\right\rangle \rightarrow\left|r\right\rangle $,
forming a Ladder-type EIT. The bandwidth of the transparency window
of Rydberg-EIT is measured as $\delta w\sim2\pi\times5$ MHz.

The probe field has a pulse width of 200 ns. The coupling field is
modulated into double rectangular pulses with a width
of 400 ns to demonstrate the write and read operations. The
amplitudes and frequencies of the write and read pulses are tuned individually
by an electro-optic modulator (EOM, LM 0202, Germany)
and an acoustic-optic modulator (AOM) respectively; therefore, we can turn
on/off the coupling field with fast rising and falling time. This
guarantees that the probe is efficiently converted into the Rydberg
polariton. We adiabatically switch off the coupling field, and a stored
\textcolor{black}{high-lying} Rydberg-state super atom is obtained
given by $1/\sqrt{N_{m}}\sum e^{i\mathbf{\mathit{\mathbf{k}}}_{S}\mathbf{\cdot\mathit{\mathbf{r}}}_{i}}\left|g\right\rangle _{1}\cdot\cdot\cdot\left|r\right\rangle _{i}\cdot\cdot\cdot\left|g\right\rangle _{N_{m}}$
\cite{fleischhauer2005electromagnetically,ding2013single,ding2015optical,ding2015raman},
also referred to as a Rydberg polariton. $\mathbf{\mathrm{\mathit{k}}}_{S}=\mathbf{\mathrm{\mathit{k}}}_{c}-\mathbf{\mathrm{\mathit{k}}}_{p}$
is the wave vector of the atomic polariton, $\mathbf{\mathrm{\mathit{k}}}_{c}$
and $\mathbf{\mathit{k}}_{p}$ are the vectors of the coupling and probe
fields and $\mathbf{\mathit{\mathbf{r}}}_{i}$ denotes the position of
the $i$-th atom in atomic cloud. After a programmed storage time,
the polariton is converted back into photonic excitation by switching
on the coupling laser again. Figure~\ref{experimental setup}(c)
shows the storage sequence for the probe pulse; the leaked and retrieved
probe fields both exhibit oscillation.

The repetition rate of our experiment is $200$~Hz, and the MOT trapping
time is 4.71~ms. Moreover, the experimental window is 290~$\mu$s.
The probe field is collected into a single-mode fiber and detected by
a single-photon detector (avalanche diode, PerkinElmer SPCM-AQR-16-FC,
60\% efficiency, maximum dark count rate of 25/s). The two detectors
are gated by an arbitrary function generator. The signal from the single-photon
detector and the trigg\textcolor{black}{ered} signal from the arbitrary
function generator are then sent to a time-correlated single-photon
counting system (TimeHarp 260) to measure the\textcolor{black}{{} probe
temporal profile. }

\begin{figure*}[t]
\includegraphics[width=1.8\columnwidth]{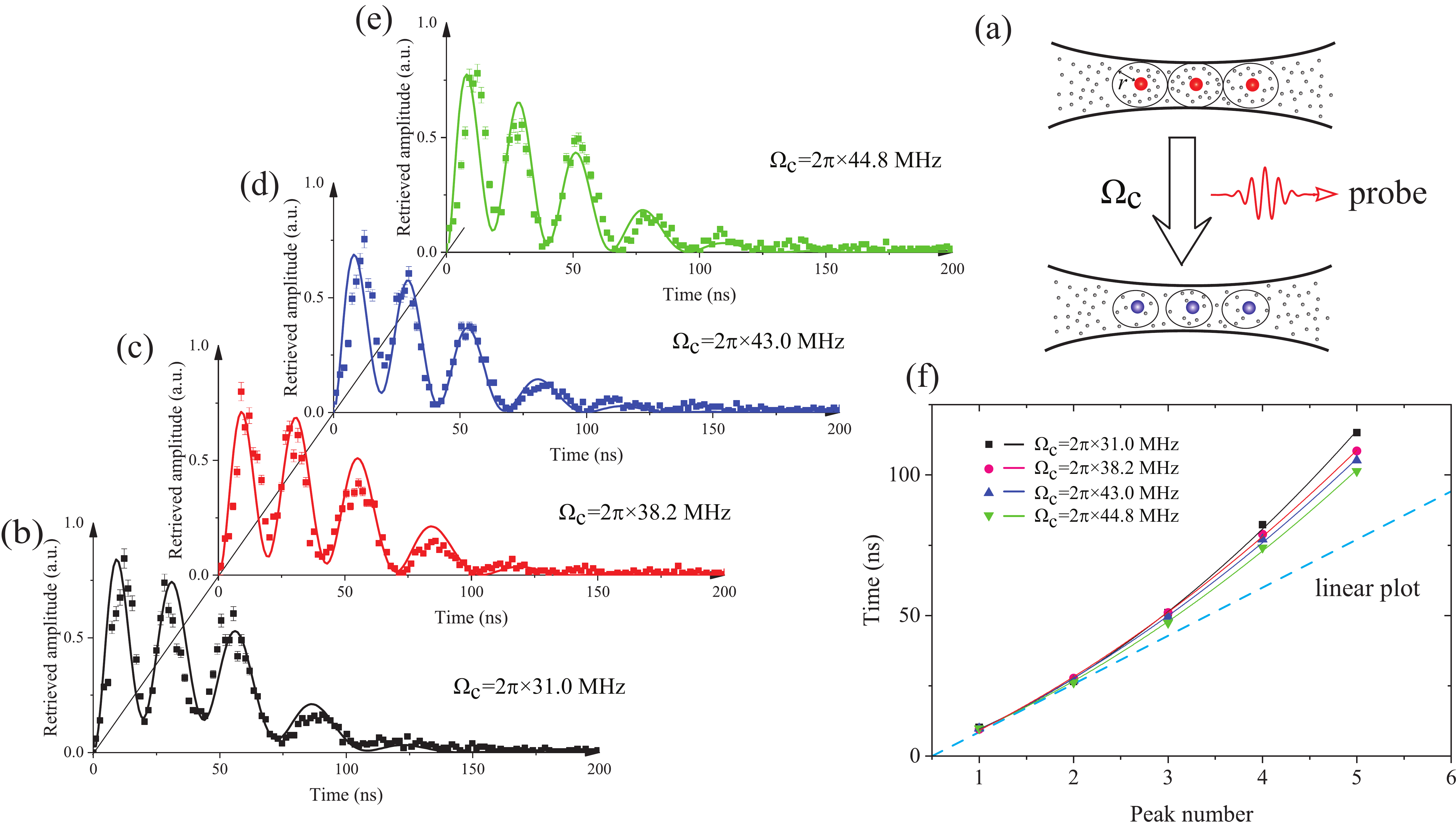}

\caption{\textbf{Coherent retrieval signal with Rabi oscillations.} (a) Schematic diagram of reading Rydberg polaritons out. The black sphere
represents the ground-state atom, the red and blue spheres correspond
to a high-lying Rydberg atom and low-lying excited atom. The size of
the collective state is reduced as the number of atoms decreases.\textcolor{black}{{}
 (b)šC(e, the retrieved probe field versus time under different
$\Omega_{c}$; solid curves are fits of the form $P_{s}(\beta,C,t_{0},\Omega_{n},t)$,
with different parameters: }$\beta$, $C$, $t_{0}$,\textcolor{black}{{},
, and $\Omega_{n}$.} (f) Measured time at different oscillation peaks. The data were fitted using the polynomial function $y=y_{0}+bx+cx^{2}$. The fitted parameters $(y_{0},b,c)$ are $(-2.7,9.5,2.8)$ for the black
data, $(-6.0,13.2,1.9)$ for red data, $(-5.1,12.6,1.92)$ for blue
data and $(-3.9,11.4,1.94)$ for green data. In this process, the detuning is $\Delta_{p}=-2\pi\times2.7$ MHz and $\Delta_{c}=2\pi\times14.8$ MHz for writing, and $\Delta_{c}=2\pi\times23.4$ MHz for reading. All error bars in the experimental data are estimated using Poisson statistics.}

\label{against power}
\end{figure*}

\textbf{\textcolor{black}{Results}}

\textbf{\textcolor{black}{Theoretical analysis}}

In the storage process, the input probe field contains approximately 10 photons
per pulse, and the efficiency of converting the photons to Rydberg polaritons
is measured as $\sim0.04$, guaranteeing one polariton excitation
in the one storage process. The probe field illuminates the entire ensemble
and excites all atoms with equal probability. Owing to the $L$-length
cylinder mesoscopic atomic ensemble along the direction of probe beam,
our system can be regarded as quasi-one-dimensional mesoscopic atomic
ensemble, see the schematic diagram in Fig. \ref{against power}(a).
After storing the probe pulse in this medium, the converted Rydberg polariton can be expressed as follows:
\begin{gather}
\left|R_{m}\right\rangle =\frac{{\rm 1}}{\sqrt{\mathcal{N}_{m}}}\sum\limits _{i=1}^{i=\mathcal{N}_{m}}e^{i\Delta\mathbf{k}\cdot\mathbf{r}_{i}}\left|g_{1}\cdots r_{i}\cdots g_{\mathcal{N}_{m}}\right\rangle
\end{gather}
where $\mathcal{N}_{m}$ is the atom number in interacted area $m$,
$\mathbf{\Delta\mathbf{k}}$ is the wave-vector mismatch between the probes,
and coupling fields, $\mathbf{r}_{i}$ is the position of atom $i$.
Accordingly, the low-lying collective excited state is given
\begin{equation}
\left|E_{m}\right\rangle =\frac{{\rm 1}}{\sqrt{\mathcal{N}_{m}}}\sum\limits _{i=1}^{i=\mathcal{N}_{m}}e^{i\mathbf{k_{p}}\cdot\mathbf{r}_{i}}\left|g_{1}\cdots e_{i}\cdots g_{\mathcal{N}_{m}}\right\rangle
\end{equation}
Owing to the atoms loss and nonlinear conversion in the reading process,
the size of the collective states $\left|E_{m}\right\rangle $ or $\left|R_{m}\right\rangle $
decreased. This reduction can be observed by observing the populations
$\left|E_{m}\right\rangle $ or $\left|R_{m}\right\rangle $ under
the driving of the coupling laser beam. The \textcolor{black}{low-lying}
collective excited state is converted into the $\mathbf{k}_{\mathbf{p}}$-photon
pulse with an efficiency of $\eta$, $\left|E_{m}\right\rangle \rightarrow\sqrt{\eta}\left|G_{m}\right\rangle +\sqrt{1-\eta}\left|E_{m}\right\rangle $,
here $\left|G_{m}\right\rangle =\left|g_{1}\cdots g_{i}\cdots g_{\mathcal{N}_{m}}\right\rangle $
which is the ground state with multiple atoms. In this process, the loss of atoms is influenced by driving the coupling
laser beam, in which there is no emitted probe field. The nonlinear
converted efficiency $\eta$ is dependent on the experimental parameters,
such as OD, $\gamma_{e}$, $\Omega_{c}$ and $\Delta_{c}$ in the
reading process, as the quantum memory is regarded as a delayed four-wave
mixing process \cite{ripka2018room}. Consequently, the collective
Rabi frequency coupling the collective state $\left|R_{m}\right\rangle $
and $\left|E_{m}\right\rangle $ becomes \cite{tresp2017rydberg}
\begin{flalign}
\Omega_{\mathrm{coll}} & =-\frac{e\varepsilon}{\hbar}\left\langle E_{m}\right|\mathbf{r}\left|R_{m}\right\rangle \\
 & =-\frac{e\varepsilon}{\hbar}\frac{{\rm \sqrt{\eta}}}{\sqrt{\mathcal{N}_{m}}}\sum\limits _{i=1}^{i=\mathcal{N}_{m^{\prime}}}\left\langle g_{1}\cdots g_{i}\cdots g_{\mathcal{N}_{m^{\prime}}}\right|\mathbf{r}\left|g_{1}\cdots r_{i}\cdots g_{\mathcal{N}_{m}}\right\rangle \nonumber \\
 & -\frac{e\varepsilon}{\hbar}\frac{{\rm \sqrt{1-\eta}}}{\mathcal{N}_{m}}\sum\limits _{i=1}^{i=\mathcal{N}_{m^{\prime}}}\left\langle g_{1}\cdots e_{i}\cdots g_{\mathcal{N}_{m^{\prime}}}\right|\mathbf{r}\left|g_{1}\cdots r_{i}\cdots g_{\mathcal{N}_{m}}\right\rangle \nonumber \\
 & = \frac{N_{m^{\prime}}}{\sqrt{N_{m}}}{\rm \sqrt{\eta}}\Omega_{g} + \frac{N_{m^{\prime}}}{N_{m}} {\rm \sqrt{1-\eta}}\Omega\nonumber
\end{flalign}
Here, $N_{m^{\prime}}$ is the remaining number of ground state atoms owing to the loss induced by blue laser driving. $\Omega_{g}=-\frac{e\varepsilon}{\hbar}\left\langle g\right|\mathbf{r}\left|r\right\rangle $,
corresponds to the Rabi frequency of a two-level single atom between the ground
state $\left|g\right\rangle $ and Rydberg state $\left|r\right\rangle $ and $\Omega=-\frac{e\varepsilon}{\hbar}\left\langle e\right|\mathbf{r}\left|r\right\rangle $
corresponds to the Rabi frequency of a two-level single atom between the excited
state $\left|e\right\rangle $ and Rydberg state $\left|r\right\rangle $.
The first term corresponds to the enhanced effective Rabi frequency, which was not considered in our experiment (we detected only the transition process between $\left|e\right\rangle $ and $\left|r\right\rangle $). The equation Eq.(3) gives rise to an anomalous oscillation between the residual low-lying collective excited state $\left|E_{m}\right\rangle $ and the high-lying Rydberg-state super atom $\left|R_{m}\right\rangle $, with a decreased Rabi frequency $ N_{m^{\prime}}\Omega{\rm \sqrt{1-\eta}}/N_{m}$. Because the quantum state $\left|E_{m}\right\rangle $ is continuously converted to $\left|G_{m}\right\rangle $ and the ground state atoms are lost during the reading process, the effective Rabi frequency $\Omega_{\mathrm{coll}}$ decreases with time during the oscillation as stated.

\begin{figure*}[t]
\includegraphics[width=1.6\columnwidth]{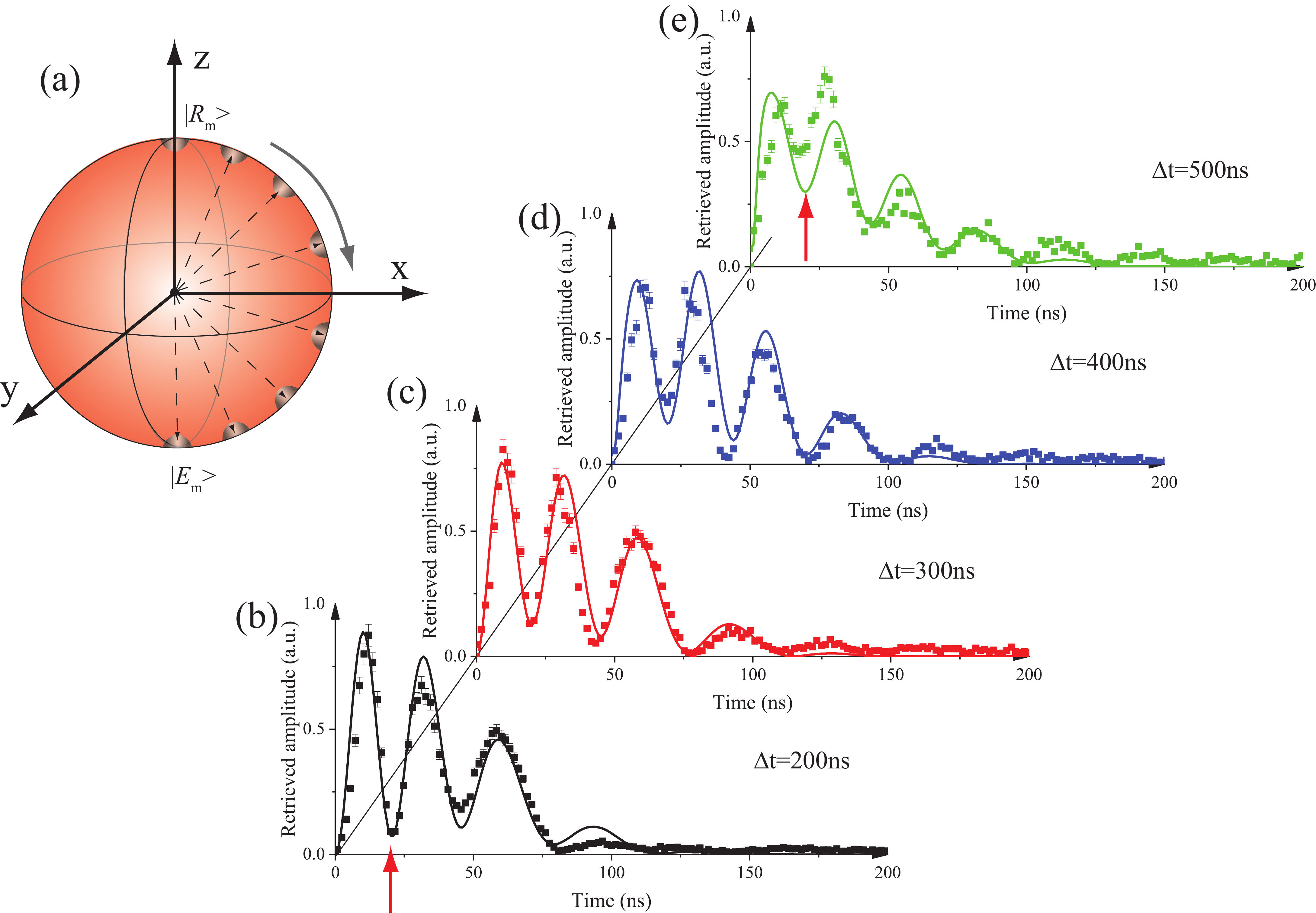}

\caption{\textbf{Measurement of retrieved probe field.} (a) Bloch sphere
for Rabi oscillation with inhomogeneous broadening. (b-e), the retrieved
probe field against time under different storage times $\Delta t$;
The solid curves are fits of the form $P_{s}(\beta,C,t_{0},\Omega_{n},t)$\textcolor{black}{.}
In this process, the detunings $\Delta_{p}=-2\pi\times2.7$ MHz and
$\Delta_{c}=2\pi\times14.8$ MHz for writing, and $\Delta_{c}=2\pi\times17.6$
MHz for reading. All error bars in the experimental
data are estimated from the Poisson statistics.}

\label{delay}
\end{figure*}

The reading process in our system can be modeled using a two-level
atomic system with a varying effective Rabi frequency because the super atom could be regarded as a quasi-single particle. In the state evolution, as we consider the inhomogeneous broadening in our system, the broadening width can be embodied by the transparency bandwidth
of the Rydberg-EIT. The probability of the retrieved signal is distributed as a Gaussian profile owing to the time reversal in the writing and reading processes \cite{novikova2007optimal,everett2018time,Ding2016}
$\sim$ $e^{-C^{2}t^{2}}$. The probability of the retrieved probe pulse under Rabi oscillation is expressed by \cite{stanojevic2009many}:
\begin{multline}
P_{r}(\beta,\Delta,C,t_{0},\Omega_{n},t)\\
=e^{-\beta^{2}(t-t_{0})^{2}}(1-\mathrm{Cos}(\sqrt{\frac{e^{-C^{2}t^{2}}+1}{2}}\Omega t))
\end{multline}
The term $e^{-\beta^{2}(t-t_{0})^{2}}$ is the fitted emission rate from a low-lying collective excited state to a photon; $C^{2}$ is the chirped coefficient, and $t_{0}$ is a parameter that fits the temporal profile of the probe intensity. The Rabi frequency $\Omega=\sqrt{\varDelta^{2}+\Omega_{n}^{2}}$.
$\Omega_{n}$ is the effective Rabi frequency of the atomic transitions $\left|r\right\rangle $ and $\left|e\right\rangle $ involving $\Omega_{c}$ and $\Delta_{c}$.
The temporal profile of the retrieved probe field can be simulated
by integrating the inhomogeneous shift $\varDelta$:
\begin{multline}
P_{s}(\beta,C,t_{0},\Omega_{n},t)\\
=\varint_{-\infty}^{+\infty}\sqrt{\pi/\alpha}e^{-\alpha\Delta^{2}}P(\beta,\Delta,C,t_{0},\Omega_{n},t)d\varDelta
\end{multline}
here, we consider the energy broadening effect, which is distributed
with a Gaussian function $\sqrt{\pi/\alpha}e^{-\alpha\varDelta^{2}}$;
here $\alpha$ is the broadening coefficient.

\textbf{Size-reduction of Rabi oscillations}

To explore the chirped character of the observed Rabi oscillations,
we measure the temporal profile of the retrieved probe field with varying
$\Omega_{c}$. Here, we set $\Delta_{p}=-2\pi\times2.7$ MHz and $\Delta_{c}=2\pi\times14.79$
MHz to write the Rydberg polariton and set $\Delta_{c}=2\pi\times23.4$
MHz to read the Rydberg polariton out. We record the retrieved probe field,
and deduce that the oscillation exhibits a period of gradual increase, which corresponds to a chirped pulse; the results are shown in Fig. \ref{against power}(b-e). The effective Rabi frequency, $\Omega_{n}$, was fitted
as $2\pi\times46.1$ MHz, $2\pi\times47.7$ MHz, $2\pi\times49.3$
MHz, and $2\pi\times50.9$ MHz, as shown in Fig. \ref{against power}(b-e), which
tend to be consistent with $\Omega_{n}^{exp}$ at large $\Omega_{c}$
by considering the effective Rabi frequency $\Omega_{n}^{exp}=\sqrt{\Omega_{c}^{2}+\varDelta_{c}^{2}}$.
The different peaks against time are plotted in Fig. \ref{against power}(f),
which are fitted by the polynomial function $y=y_{0}+bx+cx^{2}$
different from the normal fixed oscillation period with linear behavior. The collective state $\left|E_{m}\right\rangle $
is continuously converted to $\left|G_{m}\right\rangle $ during the
reading process and the collective Rabi frequency gradually decreased over time. This observation differs from previous works \cite{dudin2012observation}. The Rabi oscillation is demonstrated with fixed $N_{m}$ and $\Omega$,
thus the Rabi frequency is a constant of $\sqrt{N_{m}}$$\Omega$.

\begin{figure*}
\includegraphics[width=2\columnwidth]{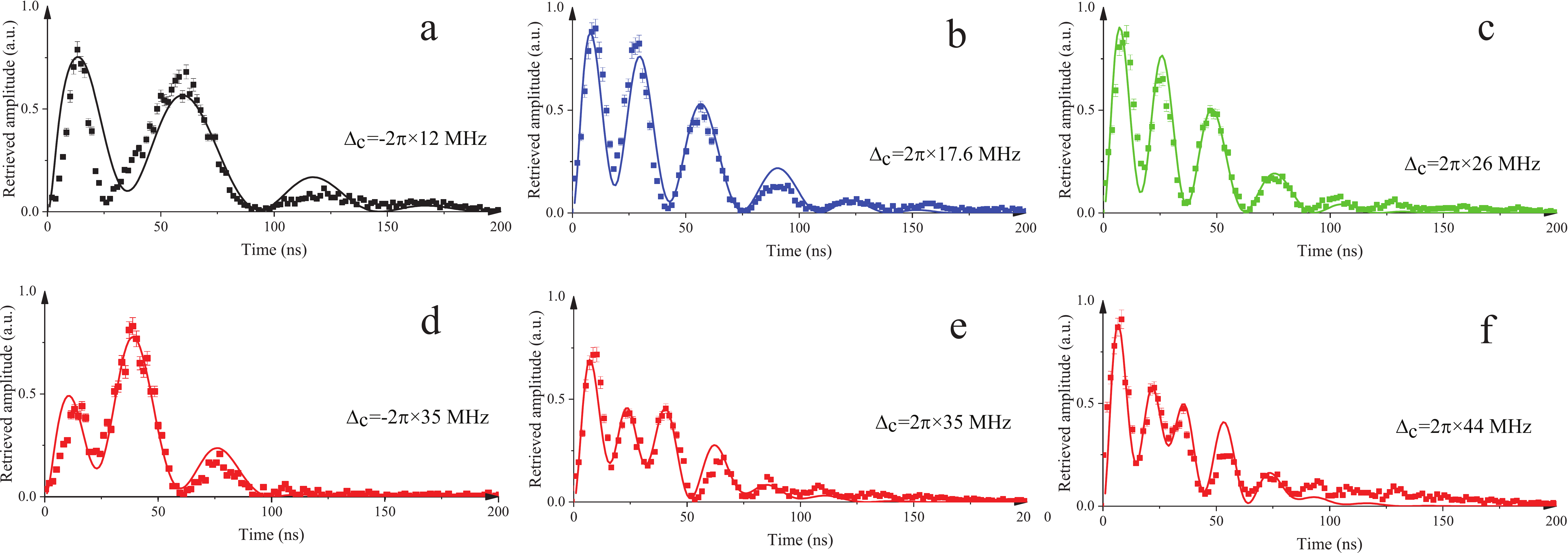}

\caption{\textbf{Measurement of retrieved probe field with different detuning $\Delta_{c}$.} (a-e) Retrieved probe field against detuning
$\Delta_{c}$; solid curves are fits of the form $P_{s}(\beta,C,t_{0},\Omega_{n},t)$ with the fit parameters $(\beta,C,t_{0},\Omega_{n})$. In this process, the storage time was set to 300 ns and $\Omega_{c}=2\pi\times44.8$
MHz. All error bars in the experimental data are estimated using Poisson statistics.}

\label{small detuning}
\end{figure*}

This process can be regarded as shaping a light pulse; the advantage of shaping a light pulse with this method is that the shaping operation is on-demand. Next, we change the storage time $\Delta t$ and record the retrieved probe field; the results are shown in Fig. \ref{delay}(b-e). Dephasing also occurs during the storage process, which affects the coherence of the Rabi oscillations between the collective $\left|R_{m}\right\rangle $ and $\left|E_{m}\right\rangle $.
In the Bloch sphere given in Fig. \ref{delay}(a), the trajectory
of rotations driven by the coupling field do not express a curve but a surface because the point on the Bloch sphere is replaced by a sphere
surface. The size of the sphere surface is determined by the broadening
coefficient $\alpha$. The broadening effect reduces the visibility
of the Rabi oscillations and may even suppress them significantly as decoherence in the storage process \cite{de2018analysis,levine2018high}.
As observed in Fig. \ref{delay}(b), the retrieved probe pulse shows an obviously decreased visibility which is marked by the red arrow. When increasing the storage time from $\Delta t=200$ ns to $\Delta t=500$ ns, the visibility is further reduced, as shown in Fig. \ref{delay}(b-e) because of the increased $\alpha$. This is because the  collective
state is dephased during storage, which generates a finite storage lifetime. The atoms in the MOT are not spin-polarized, and the absence of
spin polarization with respect to light leads to an inhomogeneous broadening of the Rabi frequencies and, therefore, to dephasing. Additionally, the inhomogeneous $\Omega_{c}$ caused by the transverse
differentiated intensity distribution of the coupling field induce
additional dephasing. The broadening effect reflects the broadened
bandwidth of the transparency window of the Rydberg EIT, as given in Ref. \cite{Yu_2020}.

Moreover, we changed the detuning $\Delta_{c}$ to explore the evolution of the collective state. In this process, we write the high-lying Rydberg-state super atom under the optimized condition $\Delta_{p}=-2\pi\times2.7$
MHz and $\Delta_{c}=2\pi\times14.8$ MHz. In the read process, the
temporal profile of the retrieved probe field is changed by varying
$\Delta_{c}$. The results are given in Fig. \ref{small detuning}(a-e),
in which the detuning $\Delta_{c}$ is changed to $-2\pi\times12$
MHz, (b) $ 2 \ pi \ times17.6 $ MHz, and (c) $2\pi\times26$ MHz (c) respectively. The periods of these oscillations are clearly
increased as the effective Rabi frequencies are increased versus $\Delta_{c}$.
The theoretical function fits the results in Fig. \ref{small detuning}(b)
and Fig. \ref{small detuning}(c) but with uncertainty deviations
for the red detuning of $\Delta_{c}$ as shown in the example in Fig. \ref{small detuning}(a).

When the detuning $\Delta_{c}$ is large enough, we can evaluate
more complex oscillations shown in Fig. \ref{small detuning}(e-f).
In these two cases, we set the detuning $\Delta_{c}=2\pi\times35$
MHz and $\Delta_{c}=2\pi\times44$ MHz, respectively. The constructive
and destructive interference appeared alternately along the time axis
in Fig.~\ref{small detuning}(e) and (f) supports a superposition
of two Rabi oscillations, with fitted Rabi frequencies $\Omega_{n}=2\pi\times41.4$,
$2\pi\times62.1$ MHz for Fig. \ref{small detuning}(e) and $\Omega_{n}=2\pi\times49.3$,
$2\pi\times68.4$ MHz for Fig.~\ref{small detuning}(f). Accordingly,
the system is described by state $\left|\psi\right\rangle =c_{1}\left|R_{m1}\right\rangle +c_{2}\left|R_{m2}\right\rangle +c_{3}\left|E_{m}\right\rangle $.
The coefficients $c_{1}$, $c_{2}$ and $c_{3}$ are the time-dependent
complex amplitudes, here $\left|R_{m1}\right\rangle $ and $\left|R_{m2}\right\rangle $
correspond to the states of high-lying Rydberg-state super atoms. However,
for the red detuning $\Delta_{c}=-2\pi\times35$ MHz in Fig. \ref{small detuning}(d)
completely opposite to the case in Fig. \ref{small detuning}(e),
there is a single oscillation with fitted Rabi frequencies $\Omega_{n}=2\pi\times41.4$
MHz. The measured data with $\Delta_{c}=\pm2\pi\times35$ MHz with
asymmetric Rabi oscillations supports that an enhanced Rabi oscillation
process occurs under blue detuning.

\textbf{\textcolor{black}{Conclusion}}

In summary, the entire process of the Rydberg-quantum
memory with Rabi oscillation can be considered as manipulating a Rydberg
super atom to shape the photon wave-packet. The unique technology to
modulate the photon wave-packet presented here is based on the Rabi oscillation
between different collective excited states. This is significantly
different from the progresses of using electro-optical modulators
to directly modulate the amplitude of single-photon wave packets \cite{kolchin2008electro,specht2009phase}
or modulating the properties of pump fields by electro-optical modulators
and spatial light modulation to change the temporal quantum waveform
of narrowband biphotons in cold atoms \cite{chen2010shaping,zhao2015shaping},
or modulating photonic bandwidth through sum frequency generation
\cite{rakher2011simultaneous,lavoie2013spectral}. The anomalous Rabi
oscillations hint that the arbitrary photonic wave-packet could be
constructed via superposing multi-polaritons with more tunable detunings.
The reported results combined the techniques of quantum memory and
the anomalous Rabi oscillations have potential in modulating the single
photon wave-packet \cite{dudin2012strongly} and provide a perspective
approach of constructing an interface between light and the atoms
to study collective effect. Additionally, this can be regarded as a
tool to realize the manipulation of the quantum state towards the
study of quantum mechanics in the microscopic field.

\textbf{Acknowledgments}

Dong-Sheng Ding and Yi-Chen Yu contributed equally to the study. The authors would like to thank Prof. Wei Zhang, Jin-Ming Cui, and Prof. Xiang-Dong.
Chen for the initial discussions on the results, Prof. Lin Li.
from Huazhong University Of Science and Technology and Prof. Yuan
Sun from the National University of Defense Technology for their valued discussions.
This study was supported by the National Key R\&D Program of China (2017YFA0304800).
the National Natural Science Foundation of China (Grant Nos. U20A20218,
61525504, 61435011), and Fundamental Research Funds for the Central
Universities, and the Youth Innovation Pro motion Association of CAS
under Grant No. 2018490.

\textbf{Conflict of interest}
The authors declare no conflict of interest.

\end{document}